\documentclass[conference]{IEEEtran}
\usepackage{amsmath}
\usepackage{eqparbox}
\usepackage{amssymb}
\usepackage{bm}
\usepackage{mathtools,mathabx}
\usepackage{amstext}

\usepackage{graphicx}
\usepackage{color}
\usepackage{booktabs}
\usepackage{longtable}
\usepackage{multicol}
\usepackage{amsfonts}
\usepackage{dsfont}
\usepackage{array}
\usepackage[most]{tcolorbox}
\usepackage[ruled]{algorithm}
\usepackage{algorithmic}
\usepackage{cases}
\usepackage{pgfplots}
\pgfplotsset{compat=newest}
\usetikzlibrary{spy}

\usepackage{subfigmat}
\usepackage{amsthm}
\usepackage{float}
\usepackage{cite}
\usepackage{epstopdf}
\def\change{black}
\newcommand{\gf}[1]{\textcolor{black}{{#1}}}

\def\pgfplotsinvokeiflessthan#1#2#3#4{%
    \pgfkeysvalueof{/pgfplots/iflessthan/.@cmd}{#1}{#2}{#3}{#4}\pgfeov
}%
\def\pgfplotsmulticmpthree#1#2#3#4#5#6\do#7#8{%
    \pgfplotsset{float <}%
    \pgfplotsinvokeiflessthan{#1}{#4}{%
        #7%
    }{%
        \pgfplotsinvokeiflessthan{#4}{#1}{%
            #8%
        }{%
            \pgfplotsset{float <}%
            \pgfplotsinvokeiflessthan{#2}{#5}{%
                #7%
            }{%
                \pgfplotsinvokeiflessthan{#5}{#2}{%
                    #8%
                }{%
                    \pgfplotsset{float <}%
                    \pgfplotsinvokeiflessthan{#3}{#6}{%
                        #7%
                    }{%
                        #8%
                    }%
                }%
            }%
        }%
    }%
}%

\theoremstyle{plain}
\newtheorem{thm}{\textbf{Theorem}}

\theoremstyle{definition}

\theoremstyle{remark}

\newcommand{\RN}[1]{%
\textup{\uppercase\expandafter{\romannumeral#1}}%
}

\usepackage{standalone}
\graphicspath{{./Figures/}}

\definecolor{azure}{rgb}{0.0, 0.5, 1.0}
\definecolor{redjigar}{rgb}{0.9, 0.01, 0.1}
\definecolor{chestnut}{rgb}{0.8, 0.36, 0.36}
\definecolor{airforceblue}{rgb}{0.36, 0.54, 0.66}
\definecolor{cadmiumorange}{rgb}{0.93, 0.53, 0.18}
\definecolor{bleudefrance}{rgb}{0.19, 0.55, 0.91}
\definecolor{carolinablue}{rgb}{0.6, 0.73, 0.89}
\definecolor{blue(ncs)}{rgb}{0.0, 0.53, 0.74}
\definecolor{dodgerblue}{rgb}{0.12, 0.56, 1.0}
\definecolor{cssgreen}{rgb}{0.0, 0.5, 0.0}
\definecolor{cadmiumgreen}{rgb}{0.0, 0.42, 0.24}
\definecolor{cadmiumorange}{rgb}{0.93, 0.53, 0.18}
\definecolor{amaranth}{rgb}{0.9, 0.17, 0.31}
\definecolor{bluegray}{rgb}{0.4, 0.6, 0.8}
\definecolor{cadmiumgreen}{rgb}{0.0, 0.42, 0.24}

\begin{document}
%
\title{Blind Asynchronous Goal-Oriented Detection for \gf{ Massive Connectivity}}

\author{\IEEEauthorblockN{Sajad Daei$^\dagger$, Saeed Razavikia$^\dagger$, Marios Kountouris$^*$, Mikael Skoglund$^\dagger$, Gabor Fodor$\dagger\flat$, Carlo Fischione$^\dagger$}
 \IEEEauthorblockA{$^\dagger$School of Electrical Engineering and Computer Science, KTH Royal Institute of Technology, Stockholm, Sweden\\
 $^*$ Communication Systems Department, EURECOM, 06410 Sophia Antipolis, France\\
$^\flat$Ericsson Research, Sweden\\ 
Email: \{sajado, sraz, skoglund, gaborf, carlofi\}@kth.se, kountour@eurecom.fr}
}

\maketitle
    
\begin{abstract}
Resource allocation and multiple access schemes are instrumental for the success of communication networks, which facilitate seamless wireless connectivity among a growing population of uncoordinated and non-synchronized users.
In this paper, we present a novel random access scheme that addresses one of the most severe barriers of current strategies to achieve massive connectivity and ultra reliable and low latency communications \gf{for 6G}. The proposed scheme utilizes wireless channels' angular continuous group-sparsity feature to provide low latency, high reliability, and massive access features in the face of limited time-bandwidth resources, asynchronous transmissions, and preamble errors. Specifically, a reconstruction-free goal oriented optimization problem is proposed which preserves the angular information of active devices and is then complemented by a clustering algorithm to assign active users to specific groups. This allows \gf{to identify} active stationary devices according to their \gf{line of sight} angles.
Additionally, for mobile devices, an alternating minimization algorithm is proposed to recover their preamble, data, and channel gains simultaneously, enabling the identification of active mobile users. 
Simulation results show that the proposed algorithm provides excellent performance and supports a massive number of devices. Moreover, the performance of the proposed scheme is independent of the total number of devices, distinguishing it from other random access schemes. 
The proposed method provides a unified solution to meet the requirements of machine-type communications and ultra reliable and low latency communications, making it an important contribution to the emerging 6G networks.
\end{abstract}

%
%
\makeatletter{\renewcommand*{\@makefnmark}{}
\footnotetext{This work was supported in part by Digital Futures. The work of M. Kountouris has received funding from the European Research Council (ERC) under the European Union’s Horizon 2020 research and innovation programme (Grant agreement No. 101003431). The work of
Gabor Fodor was supported by Digital Futures Project PERCy.

Saeed Razavikia and Carlo Fischione acknowledge the support of WASP, SSF SAICOM, VR, and Digital Futures. }
\makeatother}

%
\IEEEpeerreviewmaketitle

\section{Introduction}\label{section1}
Wireless connectivity continuously evolves and will increasingly play a critical role in people's everyday lives. Internet of Things (IoT) devices in beyond 5G (B5G) \gf{and 6G systems} will create new applications such as smart homes, smart manufacturing, autonomous vehicles, healthcare monitoring, and smart learning. To address these applications, 5G and B5G specifications have identified two essential use cases for machine-type user equipment (UE) in IoT: ultra-reliable and low latency communications (URLLC) and massive machine-type communications (mMTC). These two features will co-exist in IoT, enabling data transmission from many UEs anywhere and anytime \cite{SaeedBlind2022,mahmood2020six,razavikia2023computing}.

Random access (RA) is an indispensable component of 5G and B5G communication between \gf{UEs and base station (BS)}. There are two different ways that UEs can access the network: sourced RA (S-RA) and unsourced RA (U-RA). In S-RA, which is suitable for device-oriented applications, UEs first receive information from the network that helps them to determine when to transmit their data. In contrast, in \gf{U-RA}introduced in \cite{polyanskiy2017perspective}, the BS \gf{does not identify} the UEs, and only the transmitted messages matter for the BS. Thus, U-RA is employed in the so-called content-oriented applications.

There are \gf{two different ways} of managing data transmission in {\color{\change}  sourced RA: grant-based sourced RA (S-GBRA) and grant-free sourced RA (S-GFRA)}. In {\color{\change}S-GBRA}, UEs request permission or a grant from the BS to transmit data, while in {\color{\change}S-GBRA}, UEs transmit their data without waiting to get permission from the BS. {\color{\change} Grant-based unsourced RA is typically unnecessary due to the absence of a requirement to identify UEs in such cases. The S-GBRA} strategy~\cite{hasan2013random,bjornson2017random}, which has {\color{\change}often} been used in human-type communications, involves four stages: preamble selection, resource allocation, connection request, and contention resolution (see \cite{bjornson2017random} for more details). 
Recently, advanced methods have been proposed (see, e.g., \cite{bjornson2017random}) to reduce the collision probability {\color{\change}in S-GBRA}.
However, these types of strategies result in significant signaling overhead and latency, making them \gf{unattractive} for battery-driven \gf{mMTC} devices that have short data payloads, sparse activity rates {\color{\change}(sporadic traffic)}, and low latency requirements. To fulfill the requirements of mMTC, the GFRA scheme has been proposed in 5G as an alternative to the GB approach \cite{chen2021sparse,fengler2022pilot,xie2022massive,liu2018sparse,SajadBOD2023,daei2022Random}. GFRA allows active UEs to transmit data packets without reserving channel resources, resulting in reduced access delay. However, the number of active UEs that can access the network is severely confined by the number of available orthogonal preambles, which is limited by channel coherence time. Moreover, non-orthogonal preambles can degrade performance \cite{ke2020compressive,liu2018massive,ding2019analysis}. Polyanskiy in \cite{polyanskiy2017perspective} proposed an interesting RA scheme that does not depend on the total number of {\color{\change}inactive} devices.  The price to pay for this benefit is that users in this method are forced to select their messages from a known and shared {\color{\change}large} codebook and that the transmitted {\color{\change}(active)} messages need to be synchronized.

All in all, the theoretical study in \cite{fengler2021non} has summarized the aforementioned approaches and has shown that with the best existing approaches, {\color{\change}the number of active UEs that can access the BS in S-RA and the number of active messages that can be transmitted in U-RA are $K_a=\mathcal{O}({T^2}/{\log^2(\tfrac{K}{T^2})})$ and $K_a=\mathcal{O}({T^2}/{\log^2(\tfrac{2^B}{T^2})})$, respectively, where $K$ is the total number of users, $T$ is the random access duration (time resource) and $B$ is the number of bits that each user has to transmit in U-RA. This theoretical finding reveals that given a certain time resource $T$, the number $K_a$ of active UEs that can get a grant to access the BS in S-RA and the number of active messages that can be transmitted in U-RA depends explicitly and inversely on the total number of UEs (i.e., $K$) and messages (i.e., $2^B$), respectively, and is thus severely limited in large $K$ and $2^B$.
More explicitly, this property poses severe restrictions for massive connectivity where a very large number of UEs want to connect to the BS with sporadic and irregular traffic and can be considered as a severe barrier to cater to a very large number of IoT devices in 6G wireless networks.}


Given the aforementioned issues, the questions that arise in sourced and unsourced RA are:
\begin{enumerate}
\item {\color{\change}Does massive connectivity in sourced and unsourced RA entails massive communication resources?}
 \item How essential is it to have strong assumptions, such as synchronization and shared codebook, in \gf{U-RA?} 
 \item Is it possible to transmit a massive number of short asynchronous data payloads in U-RA?
\end{enumerate}

The succinct replies to these questions are negative, negative, and affirmative, respectively. This work provides some innovative approaches to thoroughly addressing these questions.

\input{Figures/Scheme}

\subsection{Our contributions}
This paper proposes a novel strategy called Blind asynchronous Goal-Oriented Detection (BaGOD), {\color{\change}which addresses the three questions posed in the previous section}.  BaGOD is designed based on the assumption that multiple input, multiple output (MIMO) channels exhibit angular continuous group-sparsity features. Indeed, only a few components in the angular domain contribute to the channel of each UE, and the angles of arrival (AoAs) corresponding to each UE lie alongside each other in a group, as shown in  Fig. \ref{fig.channel}. {\color{\change}BaGOD provides a novel goal-oriented optimization problem to achieve the specific goal of active user detection (AUD), which has a significantly smaller subset of information than the whole message information of all UEs. This specific optimization problem does not keep the information of UEs' messages and complex channel coefficients and is designed only to capture the goal. Interestingly, the solution to this reconstruction-free optimization problem reveals only the angle information of all active UEs. The specific feature of the BaGOD method, distinguishing it from all prior RA works, makes it independent of the total number of inactive UEs and messages in sourced and unsourced frameworks, respectively. 
After finding the angle information of active UEs by the proposed optimization problem, a clustering algorithm is developed to group active UEs based on their angles.}
 At this step, stationary active UEs are identified by their line of sight (LoS) angles. For mobile UEs, we develop an alternative minimization (AM) method to recover the preambles of mobile UEs. The identification of mobile UEs is accomplished by both their preamble and LoS angles. Mobile UEs from different sections (shown by circles in Fig. \ref{fig:schematic(a)}  allow to use of the same preambles (shown with the same color in Fig. \ref{fig:schematic(a)}. This feature allows using the same orthogonal preambles for mobile UEs located on different angular ranges (e.g., $\Theta_1, \ldots, \Theta_4$ in Fig. \ref{fig:schematic(a)} from the BS, which indeed leads to high detection performance.
 {\color{\change}The goal-oriented feature} of BaGOD provides some new features in the community of RA, which are listed below:

\textbf{Channel and noise distributions are arbitrary.} In contrast to prior RA works, such as covariance-based methods and approximate message passing (AMP) algorithms, there are no assumptions for the distributions of users' channels and noise in our method. The BS only has to know an upper bound $\eta$ for the standard deviation of the measurement noise.
    
\textbf{The total number of inactive users is irreverent.} In sharp contrast to all prior sourced RA works, our proposed strategy does not search for the active devices among a total number of devices; instead, it indirectly obtains the identity (ID) of active stationary users. However, to identify active mobile devices, an AM algorithm has to be implemented. Therefore, our work can be considered the first S-RA strategy in wireless communications that is fully independent of the total number of inactive devices. 

\textbf{Minimal communication cost.} In BaGOD, stationary devices do not require spending any resources for RA. They are identified by their LoS angles. Mobile devices in S-RA, however, require to include a preamble in their payload for their identification. In the case of U-RA, the (stationary and mobile) devices need to spend no resources for random access, and the resources are to be used only for data transmission. By incorporating angular diversity into the detection mechanism of mobile UEs, the number of orthogonal preambles required to identify mobile UEs can be substantially decreased, leading to a huge saving in resources.
    
\textbf{ Massive connectivity and ultra-reliability.} In BaGOD, communication costs at the users' side for RA are somehow transferred to the computational complexity at the BS. In fact, the number of active users that can access the BS directly depends on the computational complexity that the BS can bear and is not a theoretical limitation of BaGOD.  
In addition to providing massive connectivity for stationary UEs, BaGOD allows a massive number of  mobile UEs to access the BS by using the same orthogonal preambles for mobile UEs located at different angular ranges (see Fig. \ref{fig:schematic(a)}) with respect to the BS (as shown in Fig. \ref{fig:schematic(a)}). Besides this, the use of orthogonal preambles provides a high detection performance for mobile UEs leading to ultra-reliability. 
    
\textbf{ Unsourced RA with unknown individual codebooks.} To the best of our knowledge, all prior unsourced RA methods follow the strong assumptions of \cite{polyanskiy2017perspective}, in which the users are obliged to select one of predefined messages from a codebook that is shared among all users and BS. {\color{\change}The maximum number of active messages that can be transmitted is severely limited in large codebooks with a large number of messages.} Our method can be considered the first unsourced RA strategy in which users can send different arbitrary messages from different unknown codebooks satisfying certain assumptions in advance. {\color{\change}Moreover, the allowable number of active transmitted messages is fully independent of the total number of inactive messages.}
    
\textbf{Asynchronous and erroneous data transmission.} 
In the case of RA without any prior synchronization, there are some delays in the data transmission of UEs, which might lead to collisions, corruption, or data loss. 
Besides that, data or preamble errors in AUD can lead to inaccurate detection of active mobile users, false positives or false negatives \cite{bjornson2016random}, and inefficient use of communication resources. Unlike \cite{polyanskiy2017perspective}, the users in BaGOD do not need to be synchronized for data transmission. BaGOD takes into account these imperfect data or preamble transmission by modeling the delays and preamble errors of devices into a goal-oriented optimization and compensates for the effects of asynchronous and erroneous preambles.

This paper is a conference version of \cite{SajadBOD2023} with the following key differences:
\begin{itemize}
    \item The active UEs are asynchronous in data transmissions.
    \item There are gain-errors in the preambles of active mobile UEs, which are incorporated in the goal-oriented optimization problem.
    \item The mobile active UEs are detected based on both preambles and LoS angles as shown in Fig. \ref{fig:schematic(b)}.
\end{itemize}
\subsection{Organization of the paper} 
The paper organization is as follows: In Section \ref{sec.model}, the considered system model is presented. In Section \ref{sec.proposed}, we provide our proposed RA approach, which consists of two steps: Goal-oriented optimization \ref{sec.goal-oriented} and data/preamble recovery \ref{sec.data_rec}. Section \ref{sec.simulations} verifies the performance of our algorithm in comparison to state-of-the-art RA works by providing some numerical experiments. Lastly, the paper is concluded in Section \ref{sec.conclusion}.

\subsection{Notations}
Bold lower-case letters $\bm{x}$ indicate vector quantities, and bold upper-case $\bm{X}$ denote matrices. The transpose and Hermitian of a matrix $\bm{X}$ are represented by $\bm{X}^{\mathsf{T}}$ and $\bm{X}^{\mathsf{H}}$, respectively. For vector $\bm{x}\in\mathbb{C}^n$ and matrix $\bm{X}\in\mathbb{C}^{n_1\times n_2}$, the $\ell_2$ norm and Frobenius norm are represented by  $\|\bm{x}\|_2$, $\|\bm{X}\|_{\rm F}$, respectively. $\bm{X}\succeq \bm{0}$ means that $\bm{X}$ is a positive semidefinite matrix. For two arbitrary matrices $\bm{A}, \bm{B}$, $\langle \bm{A}, \bm{B}\rangle$ represents the trace of $\bm{B}^\mathsf{H}\bm{A}$. $\mathcal{P}_{\Omega}(\cdot)$ is an operator transforming an arbitrary matrix to a reduced matrix with rows indexed by $\Omega$. The Toeplitz matrix $\mathcal{T}(\bm{v})$ is defined as
\begin{align}\label{eq.toeplitz_mat}
	\mathcal{T}(\bm{v})=\begin{bmatrix}
		v_1&v_2&\hdots&v_{N}\\
		\overline{v}_2&v_1&\hdots&v_{N-1}\\
		\vdots&\vdots&\ddots&\vdots\\
		\overline{v}_N&\overline{v}_{N-1}&\hdots&v_1
	\end{bmatrix}
\end{align}
where the $(i,l)$-th element is given by $\mathcal{T}(\bm{v})_{(i,l)}=\left\{\begin{array}{cc}
	v_{i-l+1}&i\ge l\\
	\overline{v}_{l-i+1}& i<l
\end{array}\right\}$. $\bm{1}_{\Omega}$ is a vector of size $\mathbb{R}^N$ which has $1$s on the indices corresponding to the set $\Omega$ and zero elsewhere. ${\rm diag}(\bm{x})$ is a diagonal matrix whose main diagonal composes of the elements of $\bm{x}\in\mathbb{R}^N$.  $\bm{e}_i\in\mathbb{R}^N$ is the canonical vector whose $i$-th element is $1$ and zero elsewhere.

\section{System Model and Problem Formulation}\label{sec.model}
We consider a wireless system with a single base station (BS) equipped with a uniform linear array (ULA) composed of $N$ antennas, and $K$ user equipments (UEs), each of which has a single antenna. Among these UEs, $K_a$ are active and identified by the set $\mathcal{S}_{\rm AU}$. Following  the conventional block fading wireless channel model, wherein the channel remains constant during a coherence time $T$, the channel vector in the frequency domain, corresponding to the UE at index $k$ to the BS with $L_k\ll N$ spatial angles of arrival (AoA)\footnote{We assume there is little local scattering around BS, and the channel gains $\alpha_k(\theta)$ of user $k$ seen at the BS are constrained to lie in a small region $(\theta_k^{\min},\theta_k^{\max})$, known as an angular spread which composes of $L_k\ll N$ spatial AoA \cite{ke2020compressive,djelouat2021joint,ma2018sparse} (see Figs. \ref{fig.channel} and \ref{fig:schematic(a)}).}, can be expressed as follows \cite[Eq. 7]{zhang2017blind}:
\begin{align}\label{eq.chann}
	\bm{h}_k=\sum\nolimits_{l=1}^{L_k}\alpha_l^k\bm{a}(\theta_l^k)=:\bm{A}_k\bm{\alpha}^k\in\mathbb{C}^{N\times 1}.
\end{align}
where $\alpha_l^k$ accounts for the gain of the $l$-th path,
$\theta_l^k$ is the AoA of the $l$-th path for the $k$-th user and $\bm{\alpha}^k:=[\alpha_1^k,..., \alpha_{L_k}^k]^\mathsf{T}$, $\bm{A}_k:=[\bm{a}_r(\theta_1^k),..., \bm{a}_r(\theta_{L_k}^k)]\in\mathbb{C}^{N\times L_k}$,
and $\bm{a}(\theta)$ is the array response vector of BS defined by
\begin{align}\label{eq.atoms}
	\bm{a}(\theta)=\tfrac{1}{\sqrt{N}}[1, {\rm e}^{-j2\pi \frac{d}{\lambda} \cos(\theta)},...,{\rm e}^{-j2\pi \frac{d}{\lambda} (N-1)\cos(\theta)} ]^\mathsf{T},
\end{align}
where $\lambda$ and $d$ are the carrier wavelength and antenna spacing, respectively. 
Then, the BS uses only the received signals by $M$ antennas over $N$ indexed by $\Omega \subseteq \{1,..., N\}$ ($|\Omega|=M$) for each UE. Let  $\bm{\phi}_k\in\mathbb{C}^{T\times 1}$ be the information transmitted by $k$-th user\footnote{For simplicity, the data length of all UEs are considered to be the same.}. Consider a practical case where active users have time-lag delays $\tau_{e_k}, k=1,..., K_a$ with maximum delay $\tau_{\max}$\footnote{Alternatively, their corresponding channels can have different delays.}, and their data have been under some gain errors, which are denoted by $\bm{g}_k\in \mathbb{R}^{T\times 1}, k=1, ..., K_a$.
\input{Figures/Fig_Channel}
The received signal at the BS in the frequency domain becomes in the form of \cite[Eq. 7]{liu2018massive}, \cite[Eq. 5]{SajadBOD2023}:
\begin{align}\label{eq.observed}
	\bm{Y}_{\Omega}& = 
 \mathcal{P}_{\Omega}\big(\sum_{k\in\mathcal{S}_{\rm AU}}\bm{h}_{k}\widetilde{\bm{\phi}}_k^{\mathsf{H}}\bm{F}_{\rm ext}^{\mathsf{H}}\big)+\hspace{-2pt}\bm{N}= \hspace{-6pt}\sum_{k\in\mathcal{S}_{\rm AU}}\mathcal{P}_{\Omega}(\bm{h}_k\bm{\varphi}_k^{\mathsf{H}}\bm{E}_k)\hspace{-1pt} +\hspace{-1pt}\bm{N} \nonumber\\&:=\sum\nolimits_{k\in\mathcal{S}_{\rm AU}}  \mathcal{P}_{\Omega}(\bm{X}_k)+\bm{N}\in \mathbb{C}^{M\times T},
\end{align}  
where $\widetilde{\bm{\phi}}_k\in\mathbb{C}^{(T+\tau_{\max})\time 1}$ is the zero-padded, delayed version of $\bm{\phi}_k$, $\bm{F}_{\rm ext}\in\mathbb{C}^{T\times (T+\tau_{\max})}$ is the partial discrete Fourier transform (DFT) matrix
, $\bm{\varphi}_k:= \bm{F}_k \bm{\phi}_k\in\mathbb{C}^{T\times 1}$, $\bm{F}_k:=\bm{P}_k \bm{F}\in\mathbb{C}^{T\times T}$ is the DFT matrix applied to the last $T$ samples of $\widetilde{\bm{\phi}}_k$, $\bm{F}\in\mathbb{C}^{T\times T}$ is the regular DFT matrix of size $T$ and $\bm{P}_k$ is the permutation matrix. Also, $\bm{X}_k:=\bm{h}_{k}\bm{\varphi}_k^{\mathsf{H}}\bm{E}_k$, ,  $\bm{E}_k = {\rm diag}( \widetilde{\bm{e}}_k) \in \mathbb{C}^{T\times T}$, and $\widetilde{\bm{e}}_k = [1, {\rm e}^{j2\pi \tau_{e_k}\frac{1}{T}},\ldots,{\rm e}^{j2\pi \tau_{e_k}\frac{T-1}{T}}]^{\mathsf{T}}\odot (\bm{1}_{T\times 1}+\bm{g}_k)$ and $\tau_{e_k}$ denotes the time-delay of UE $k$. Note that $\bm{E}_k$s are delay-gain error matrices that model the continuous-valued unknown time-lag delays between UEs and gain error of the preambles. Also, 
 $\bm{N}\in\mathbb{C}^{M\times T}$ denotes the additive noise matrix which has arbitrary distribution with $\|\bm{N}\|_{\rm F}\le \eta$ and $\mathcal{S}_{\rm AU}\subseteq \{1,\ldots, K\}$ is the set of active users. $\bm{X}_k$ can be described as a sparse linear combination of some building blocks (so-called atoms) taken from the atomic set 
 $\widetilde{\mathcal{A}}_k:=\{\bm{a}(\theta_l^k)\bm{\varphi}_k^\mathsf{H}\bm{E}_k: \|\bm{\varphi}_k\|_2=1, \theta\in (0,\pi), \|\bm{E}_k\|_{2\rightarrow 2}\le C_e\}$ where $C_e:=1+\zeta$ and $\zeta$ is an upper-bound for the gain error, i.e., $\max_k \|\bm{g}_k\|_{\infty}\le \zeta$.

\section{Proposed method}\label{sec.proposed}

In this section, we describe our blind RA strategy. First, in Subsection \ref{sec.goal-oriented}, we design a goal-oriented optimization problem that only reflects the ID of active UEs without doing any reconstruction tasks. In the second step provided in Subsection \ref{sec.data_rec}, data recovery, channel estimation, and gain-error of preambles are investigated. It is worth mentioning that the first step is sufficient to identify active stationary UEs by their LoS angles, while for mobile active UEs, preamble and LoS angles are both required (see Fig. \ref{fig:schematic(b)}). A summary of these two steps is provided in Algorithm \ref{algorithm.admm}.

\subsection{Goal-oriented optimization problem} \label{sec.goal-oriented}  

While the under-determined system of equations in \eqref{eq.observed}
 has $K N T$ number of unknowns and only $M T$ observations at the BS, the degrees of freedom thereof is $\sum_{k\in\mathcal{S}_{\rm AU}}(L_k+T)$. This motivates us to use a challenging optimization problem that promotes angular sparsity features of all the channels corresponding to all UEs as follows: (see \cite{sayyari2020blind,Seidi2022Novel,SaeedBinary2020,daei2019living,daei2019error,daei2019exploiting,maskan2023demixing,valiulahi2019two,candes2014towards,tang2013compressed,fernandez2016super} for more details):
\begin{align}\label{prob.atomic_l1_lasso}
	&\min_{\substack{\bm{Z}_k\in\mathbb{C}^{N\times T}\\ k=1,..., K\\\bm{Y}^{\star}\in\mathbb{C}^{M\times T}}} \sum_{k=1}^K \beta_k\|{\bm{Z}_k}\|_{\widetilde{\mathcal{A}}_k}+\frac{\gamma}{2}\|\bm{Y}-\bm{Y}^{\star}\|_{\rm F}^2 \nonumber\\
	&~s.t.~~ \bm{Y}^{\star}=\sum\nolimits_{k=1}^K\mathcal{P}_{\Omega}(\bm{Z}_k),
\end{align}
where $\gamma>0$ is a regularization parameter that makes a balance between the noise energy and angular sparsity feature, and the atomic norm $\|\cdot\|_{\mathcal{A}_k}$ is the best convex surrogate for the number of atoms composing $\bm{Z}_k$ which is defined as the minimum of the $\ell_1$ norm of the coefficients forming $\bm{Z}_k$:
\begin{align}
\label{prob.atomic_l0_lasso}
	&\|\bm{Z}_k\|_{\widetilde{\mathcal{A}}_k}:=\inf\{t>0: \bm{Z}_k\in t{\rm conv}(\widetilde{\mathcal{A}}_k)\} = \inf\{\sum_{l=1}^{L_k}c_l^k: \bm{Z}_k\nonumber\\
 &=\sum_{l=1}^{L_k}c_{l}^k\bm{a}({\theta}_l^k)\bm{\varphi}_k^\mathsf{H}\bm{E}_k,~c_l^k>0, \|\bm{E}_k\|_{2\rightarrow 2}\leq C_e, \|\bm{\varphi}_k\|_2 = 1\},
\end{align}
where ${\rm conv}(\widetilde{\mathcal{A}})$ is the convex hull of $\widetilde{\mathcal{A}}$.
$\beta_k$s are some parameters that are employed to compensate for the effects of synchronization and gain errors. {\color{\change}If $\beta_k=1$, then  $\sum_{k=1}^K \|{\bm{Z}_k}\|_{\widetilde{\mathcal{A}}_k}$ is a convex surrogate for the total number of paths, i.e., $\sum_{k=1}^K L_k$.} 
{\color{\change}While the optimization problem stated in \eqref{prob.atomic_l0_lasso} may initially appear manageable due to its convexity, it presents significant challenges due to the continuous nature of the angles $\theta_k$. This continuous aspect renders the problem computationally intractable. Additionally, \eqref{prob.atomic_l0_lasso} is designed to extract both message and channel information for all UEs. However, the primary objective of this section is to detect the AoAs of active users, which represents a considerably smaller subset of the complete message and channel information for all users.
To address this objective, we propose a new theorem that introduces a reconstruction-free optimization problem that only aims to reflect the goal of active user detection (AUD) (a goal-oriented feature related to the AoA information specifically corresponding to active users.)}


\begin{thm}\label{thm.main}
    Let 
    $c_1:=C_e\frac{\max_{k=1,..., K}\|\bm{\phi}_k\|_2}{\min_{k} \beta_k\sqrt{N}}$ and $\bm{V}^*$ be the solution to the following optimization problem:
	\begin{equation}\label{prob.goal_optimization}
        \begin{aligned}
		&\bm{V}^* = \underset{\substack{\bm{V}\in\mathbb{C}^{N\times T}\\ \bm{Q}\in \mathbb{C}^{N\times N}}}{\rm argmax}~~{\rm Re}\langle \bm{V}, \bm{Y}\rangle-\frac{1}{2\gamma}\|\bm{V}\|_{\rm F}^2\\
		&~~~~{\rm s.t.}~~
		\begin{bmatrix}
			\bm{Q}&\mathcal{P}^{{\rm Adj}}_{\Omega}(\bm{V}) c_1\\
			(\mathcal{P}^{{\rm Adj}}_{\Omega}(\bm{V}))^{\mathsf{H}} c_1&\bm{I}_T
		\end{bmatrix}\succeq \bm{0},\\
		& ~~~~~~\langle \mathcal{T}({\bm{e}_q}), \bm{Q}  \rangle=1_{q=0}, \quad q=-N+1,..., N-1.
        \end{aligned}
	\end{equation}	
Then, the AoAs corresponding to the active users are uniquely identified provided that $\Delta> \frac{1}{N}$, where
 $\Delta:=\min_{k=1,..., K}\min_{i\neq q}|\cos(\theta_i^k)-\cos(\theta_q^k)|,$\footnote{ The absolute value only for minimum separation $\Delta$ is evaluated over the unit circle, e.g., $|0.2-0.8|=0.4$.}
 by finding angles that maximize the $\ell_2$ norm of the goal-oriented dual polynomial $\bm{q}_G(\theta)$ as follows:
	\begin{align}\label{eq.angle_find}
		\widehat{\theta}_l^k=\mathop{\arg\max}_{\theta\in (0,\pi)} \|\bm{q}_G(\theta)\|_2,~~l=1,..., L_k, k\in\mathcal{S}_{\rm AU},
	\end{align}
	where $\bm{q}_G(\theta):=(\mathcal{P}_{\Omega}^{{\rm Adj}}(\bm{V}^*))^\mathsf{H}\bm{a}(\theta)$.
\end{thm}
Proof. See Appendix \ref{Sec.Proof}.

\begin{figure}[!t]
\centering
 \begin{tikzpicture} 
    \begin{axis}[
        title = {\footnotesize $K=100, K_a =3, L_{\rm max} = 3, N=32, T =1$},
        width=9cm,
        height=7cm,
        xmin=0, xmax=3.14,
        ymin=3e-3, ymax=1.2,
        legend style={nodes={scale=0.7, transform shape}, at={(0.97,0.95)}}, 
        ticklabel style = {font=\footnotesize},
        ymajorgrids=true,
        xmajorgrids=true,
        grid style=dashed,
        grid=both,
        grid style={line width=.1pt, draw=gray!10},
        major grid style={line width=.2pt,draw=gray!30},
    ]
    \addplot[ smooth,
             thin,
        color=airforceblue,
        line width=1pt,
        ]
    table[x=x1,y=y1]
    {Data/Poly2.dat};
    \addplot[ 
        color=cadmiumorange,
        mark=star,
        dashed,
        line width=1pt,
        mark size=2pt,
        ]
    table[x=x1,y=y2]
    {Data/Poly2.dat};
    \addplot[ 
        only marks = True,
        color=redjigar,
        mark=o,
        mark size=1.5pt,
        ]
    table[x=x1,y=y2]
    {Data/Poly2.dat};
    \legend{$\|\bm{q}_{G}(\theta)\|_2$, Estimated angles, Original angles};
    \end{axis}
    \node[above,red] at (1.4,5) {\scriptsize AoAs of active users};
    \draw (0.65,4.5) circle [black, radius=0.2];
    \draw[->,>=stealth](1.1,5)--(0.9,4.8);
    \draw (2.1,4.5) circle [black, radius=0.2];
    \draw[->,>=stealth](1.8,5)--(2,4.8);
    \draw (4.15,4.5) circle [black, radius=0.23];
    \draw[->,>=stealth](2.3,5)--(3.6,4.75);
\end{tikzpicture}

  \caption{The $\ell_2$ norm of the goal-oriented dual polynomial function i.e. $\|\bm{q}_{G}(\theta)\|_2$. The angles for which $\|\bm{q}_{G}(\theta)\|_2$ achieves its maximum determine the angles of active UEs }
   \label{fig.dualpol}
\end{figure}
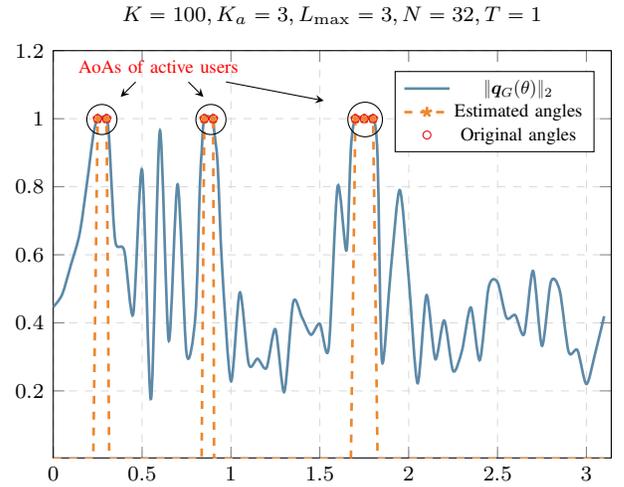

The goal-oriented dual polynomial $\bm{q}_G(\theta)$ contains all the necessary information for identifying active users. A typical example of the $\ell_2$ norm of this function, $\|\bm{q}_G(\theta)\|_2$, is shown in Fig. \ref{fig.dualpol}. The angles corresponding to active devices can be obtained by maximizing $\|\bm{q}_G(\theta)\|_2$. The proposed optimization problem \eqref{prob.goal_optimization} is agnostic to the number of devices, $K$, and can be solved using optimization solvers, such as CVX~\cite{grant2014cvx}. However, solving this problem can be challenging in massive MIMO communication scenarios where $N\gg1$, and conventional solvers, such as the SDPT3 solver of the CVX package, may not be able to provide the solution. To address this issue, a fast ADMM method can be designed, which reduces the computational complexity tailored to massive access~\cite[Algorithm 1]{SajadBOD2023}. Obtaining $|\bm{q}_G(\theta)|_2$ requires the matrix dual solution $\bm{V}$, which is typically returned along with the primal variables by most CVX solvers.

Optimization Problem \eqref{prob.goal_optimization} yields $\mathcal{P}^{{\rm Adj}}(\bm{V})$, which can be utilized alongside \eqref{eq.angle_find} in Theorem \ref{thm.main} to estimate active angles. However, the association of angles with corresponding devices is not straightforward. We design a clustering method that groups angles into several clusters that correspond to active devices.
First, we discretize the continuous function $q_G(\theta)$ on a desired angular grid. Second, we form the vector of angles achieving the maximum within a specific range. Third, we identify the boundary of clusters (jump points) by realizing the large variations in the vector of angles. By this task, all of the clusters with their lengths can be identified. The number of clusters denoted by $\widehat{K}_a$ provides an estimate of $K_a$, and the number of angles within the $k$-th cluster denoted by $\widehat{L}_k$ gives an estimate of $\widehat{L}_k$. As a result, we can find estimates for the angles of each active user denoted by $\{\widehat{\bm{\theta}}^k\}_{k=1}^{\widehat{K}_a}$.
At this step, the BS can directly identify active stationary devices by their LoS angles. To identify active mobile devices, orthogonal preambles are also exploited, as detailed in the following subsection.

\subsection{Recovery of preambles, channel amplitudes and delay-gain matrices}\label{sec.data_rec}

We propose an alternative minimization scheme to retrieve the preamble of each active device and the delay-gain errors. In particular, the following optimization problem should be solved:
\begin{align}\label{eq.alternative_opt1}
	&\mathop{\min}_{\substack{\bm{\alpha}^k:=[\alpha_1^k,...,\alpha_{L_k}^k],\\ \bm{E}_k,\bm{P}_k, \bm{\phi}_k}}
\|\bm{Y}-\sum_{k\in\widehat{\mathcal{S}}_{\rm AU}}\sum_{l=1}^{\widehat{L}_k}\alpha_l^k\mathcal{P}_{\Omega}(\bm{a}(\widehat{\theta}_l^k)\bm{\phi}_k^\mathsf{H} \bm{F}^\mathsf{H} \bm{P}_k^{\mathsf{H}} \bm{E}_k)\|_{\rm F}\nonumber\\
&\text{s.t.}~~ \|\bm{E}_k\|_{2\rightarrow 2}\le C_e, \bm{\phi}_k\ge \bm{0}, \|\bm{\phi}_k\|_2=1, k=1, \ldots, \widehat{K}_a,
\end{align}
where the non-negative constraint $\bm{\phi}_k\ge \bm{0}$ is included to remove the multiplication ambiguities of multiple elements. Due to the unitary properties of the permutation matrices, the latter optimization problem can be further simplified to
\begin{align}\label{eq.alternative_opt}
	&\mathop{\min}_{\substack{\bm{\alpha}^k:=[\alpha_1^k,...,\alpha_{L_k}^k], \\\bm{E}_k, \bm{\phi}_k}}
\|\bm{Y}-\sum_{k\in\widehat{\mathcal{S}}_{\rm AU}}\sum_{l=1}^{\widehat{L}_k}\alpha_l^k\mathcal{P}_{\Omega}(\bm{a}(\widehat{\theta}_l^k)\bm{\phi}_k^\mathsf{H} \bm{F}^\mathsf{H} \bm{E}_k)\|_{\rm F}\nonumber\\
&\text{s.t.}~~ \|\bm{E}_k\|_{2\rightarrow 2}\le C_e, \bm{\phi}_k\ge \bm{0}, \|\bm{\phi}_k\|_2=1, k=1, \ldots, \widehat{K}_a,
\end{align}
where $\bm{P}_k$ is integrated into diagonal matrix $\bm{E}_k$ without changing the spectral norm constrains.
By using an alternative numerical optimization, the estimates of complex channel amplitudes, preambles, and permuted delay-gain error matrix, denoted by
$\widehat{\bm{\alpha}} := [\widehat{\bm{\alpha}}^1,..., \widehat{\bm{\alpha}}^{\widehat{K}_a}]$, $\bm{\Phi}:=[\bm{\phi}_1,...,\bm{\phi}_{\widehat{K}_a}]\in\mathbb{C}^{T \times \widehat{K}_a}$ and $\widehat{\bm{E}}_k$, respectively, can be recovered.

Following the successful recovery of the users' preamble $\bm{\phi}_k$s, the user IDs and LoS angles are both exploited to identify active mobile UEs.
The pseudo-code of the proposed method, which is indeed a summary of the aforementioned steps, is provided in Algorithm \ref{algorithm.admm}.
\begin{algorithm}[h]
        \caption{Blind asynchronous Goal-Oriented Detection (BaGOD)}
        \begin{algorithmic}[1]\label{algorithm.admm}
            \REQUIRE $\bm{Y}\in\mathbb{C}^{M\times T}$
            \STATE Obtain $\bm{V}^*$ by solving optimization in \eqref{prob.goal_optimization}.
            \STATE Compute $\bm{q}_{G}(\theta) = (\mathcal{P}_{\Omega}^{{\rm Adj}}(\bm{V}^*))^\mathsf{H}\bm{a}(\theta)$ by discretizing $\theta$ on a fine grid up to a desired accuracy. 
            \STATE Localize the angles by finding all the maximums of $\ell_2$ norm of $\bm{q}_G(\theta)$,i.e.,\\ $\widehat{\theta}_l^k=\mathop{\arg\max}_{\theta\in (0,\pi)} \|\bm{q}_G(\theta)\|_2,~~l=1,..., L_k, k\in\mathcal{S}_{\rm AU},$
            \STATE \text{Perform the clustering method:} $[\{\widehat{\bm{\theta}}^k\}_{k\in \widehat{\mathcal{S}}_{\rm AU}},\widehat{K}_a,\{\widehat{L}_k\}_{k\in \widehat{\mathcal{S}}_{\rm AU}}]={\rm cluster}(\bm{\widehat{\theta}})$.
            \STATE Solve the optimization problem \ref{eq.alternative_opt} to obtain the estimates $\widehat{\bm{\alpha}}^k$, $\widehat{\bm{\phi}}_k$ and delay-gain matrix $\widehat{\bm{E}}_k$
        \end{algorithmic} 
        Return: $\widehat{\mathcal{S}}_{\rm AU}, \{\widehat{\bm{\theta}^k}\}_{ k\in\widehat{\mathcal{S}}_{\rm AU}},\{\widehat{\bm{\phi}_k}\}_{ k\in\widehat{\mathcal{S}}_{\rm AU}}, \{\widehat{\bm{\alpha}^k}\}_{ k\in\widehat{\mathcal{S}}_{\rm AU}}, \widehat{K}_a= |\widehat{\mathcal{S}}_{\rm AU}|, \{\widehat{L}_k\}_{ k\in\widehat{\mathcal{S}}_{\rm AU}}, \widehat{\bm{h}}_k:=\sum_{l=1}^{\widehat{L}_k}\widehat{\alpha}_l^k\bm{a}(\widehat{\theta}_l^k),\widehat{\bm{E}}_k, ~ k\in\widehat{\mathcal{S}}_{\rm AU} $.
    \end{algorithm}

\section{Simulations}\label{sec.simulations}
In this section, we evaluate the detection performance of the BaGOD method and compare it to the AMP approach employed in \cite{ke2020compressive}\footnote{It should be noted that, unlike \cite{SajadBOD2023}, covariance-based methods are not utilized in the numerical results due to their inability to detect in the specific setting under consideration effectively.}. AMP uses the resource $T$ for AUD and channel estimation, while BaGOD uses this resource for data recovery of stationary UEs and preamble and data recovery of mobile UEs. Here, we employ Gaussian distributed $\bm{\phi}_k$s for AMP known to be the best distribution for AMP as discussed in \cite{ke2020compressive} and a positive $\bm{\phi}_k$ uniformly distributed from $U(0,1)$ for BaGOD method. We consider different delays $\tau_k$s between $\bm{\phi}_k$s of active UEs, which alternatively represent different phases of $\bm{\varphi}_k$s in the frequency domain. Moreover, we have considered some gain-error in the preambles of active mobile users, which affects the preamble recovery in Subsection \ref{sec.data_rec}.

\input{Figures/Fig_Sim}

The RA performance is quantified in terms of detection and false alarm probabilities, defined by $\mathds{P}_d=\frac{\mathds{E}|\mathcal{S}_{\rm AU}\bigcap \widehat{\mathcal{S}}_{\rm AU}|}{K_a}$ and $\mathds{P}_{fa}=\frac{\mathds{E}|\widehat{\mathcal{S}}_{\rm AU} \setminus \mathcal{S}_{\rm AU}|}{K-K_a}$, respectively, where $|\widehat{\mathcal{S}}_{\rm AU} \setminus \mathcal{S}_{\rm AU}|$ counts the number of elements in $\widehat{\mathcal{S}}_{\rm AU}$ that are not present in $\mathcal{S}_{\rm AU}$. Moreover, the detection and false alarm probabilities regarding stationary and mobile users are defined as $\mathds{P}_{d,{\rm S}}=\frac{\mathds{E}|\mathcal{S}_{\rm SAU}\bigcap \widehat{\mathcal{S}}_{\rm SAU}|}{K_{a,{\rm S}}}$, $\mathds{P}_{fa,{\rm S}}=\frac{\mathds{E}|\widehat{\mathcal{S}}_{\rm SAU} \setminus \mathcal{S}_{\rm SAU}|}{K_{\rm S}-K_{a,{\rm S}}}$, and $\mathds{P}_{d,{\rm M}}=\frac{\mathds{E}|\mathcal{S}_{\rm MAU}\bigcap \widehat{\mathcal{S}}_{\rm MAU}|}{K_{a,{\rm M}}}$, $\mathds{P}_{fa,{\rm M}}=\frac{\mathds{E}|\widehat{\mathcal{S}}_{\rm MAU} \setminus \mathcal{S}_{\rm MAU}|}{K_{\rm M}-K_{a,{\rm M}}}$, respectively.

The simulations are done for $50$ Monte-Carlo trials to calculate  $\mathds{P}_{d}, \mathds{P}_{fa}$. The regularization parameter $\gamma$ is set to $\frac{1}{\eta}$, where $\eta$ is an upper bound for the standard deviation of the measurement noise, i.e., $\eta\ge \|\bm{N}\|_{\rm F}$. The signal to noise ratio is defined by ${\rm SNR}=10\log_{10}(\frac{\|\mathcal{P}_{\Omega}(\bm{X})\|_{\rm F}^2}{MT\sigma^2})$. Fig.~\ref{fig:Sim} shows  comparisons between BaGOD and AMP methods with different $T, N, K_S, K_M, K_{a, M}, K_{a, S}$. Here, $K_S$ and $K_M$ represent the total number of stationary and mobile UEs, while $K_{a,M}$ and $K_{a,S}$ indicate the number of active mobile and stationary UEs, respectively.  As it turns out, BaGOD provides a substantial performance improvement in the detection of both stationary and mobile devices in the considered scenarios. When the number of active UEs increases, the number of peaks in the goal-oriented dual polynomial (as illustrated in Fig.~\ref{fig.dualpol}) also increases. This, in turn, makes it more difficult to identify the active UEs and ultimately results in a decrease in detection performance, as demonstrated in Fig.~\ref{fig:sim(a)} and Fig.~\ref{fig:sim(b)}. $K_S$ does not have any role in the detection, as shown in Fig.~\ref{fig:sim(e)}. The utilization of LoS angles and orthogonal preambles for detecting active mobile UEs has resulted in a minimal impact of $K_M$ on detection performance, as evidenced by the data presented in Fig.~\ref{fig:sim(c)}. As depicted in Fig.~\ref{fig:sim(f)}, an increase in the number $N$ of antennas results in sharper peaks of active UEs, making identification easier and improving detection performance. Fig.~\ref{fig:sim(d)} shows that an increase in the number $T$ of orthogonal preambles improves the identification of active mobile UEs. Moreover, the probability of detection and false alarm of AMP both tend towards one as $T$ increases.

Furthermore, we observe that asynchronous active UEs can be identified by BaGOD perfectly, and asynchronous UEs do not make RA harder as long as the maximum delays do not exceed the channel coherence time. Aside from this, if we know an upper bound for the maximum preamble gain distortion, i.e., $\zeta\ge \max_k \|\bm{g}_k\|_{\infty}$, we can compensate the effect of gain error in preamble recovery by solving \eqref{eq.alternative_opt} which plays a key role in active mobile UE detection.

\section{Conclusion}\label{sec.conclusion}

This paper proposes a novel strategy for RA in B5G and 6G networks based on reconstruction-free optimization. A goal-oriented optimization approach is introduced to find a valuable continuous function that provides sufficient information to obtain the ID of active users. The proposed method achieves the goal of active user detection without reconstructing the corresponding channels and messages, making it independent of the number of devices. The proposed
scheme offers a promising solution to overcome the challenges of massive MTC and URLLC in IoT applications, which \gf{are} essential for the success of smart cities and
connected vehicles. Simulation results show that the proposed method meets the URLLC and massive connectivity requirements of 6G networks and \gf{outperforms} existing RA methods.


\appendices
\section{Proof of Theorem \ref{thm.main}}\label{Sec.Proof}

By following similar steps as in \cite[Appendix B]{SajadBOD2023}, one can obtain the dual problem of \eqref{prob.atomic_l1_lasso} as follows
\begin{align}\label{d4}
	&\max_{\bm{V}\in\mathbb{C}^{N\times T}}{\rm Re}\langle \bm{V}, \bm{Y}\rangle-\frac{1}{2\gamma}\|\bm{V}\|_{\rm F}^2\nonumber\\
	&{\rm s.t.}~~\|\mathcal{P}^{{\rm Adj}}_{\Omega}(\bm{V})\|_{\widetilde{\mathcal{A}}_k}^{\mathsf{d}}\le \beta_k,~\forall \theta_k \in (0,\pi),~k=1,..., K.
\end{align}
Here, $\|\cdot\|^d_{\widetilde{\mathcal{A}}_k}$ is the dual norm associated with the atomic norm $\|\cdot\|_{\widetilde{\mathcal{A}}_k}$ and is defined as
\begin{align}
&\|\mathcal{P}^{{\rm Adj}}_{\Omega}(\bm{V})\|^d_{\widetilde{\mathcal{A}}_k}:=\sup_{\|\bm{Z}\|_{\widetilde{\mathcal{A}}_k}\le 1}\langle \mathcal{P}^{{\rm Adj}}_{\Omega}(\bm{V}), \bm{Z}\rangle.
\end{align}
Recall the definition of $\|\cdot\|_{\widetilde{\mathcal{A}}_k}$ from \eqref{prob.atomic_l0_lasso}, we can write
\begin{align}
\|\mathcal{P}^{{\rm Adj}}_{\Omega}&(\bm{V})\|^d_{\widetilde{\mathcal{A}}_k} = \sup_{\substack{\theta_k\in (0,\pi),\bm{\varphi}_k,\\\|\bm{E}_k\|_{2\rightarrow 2}\le C_e}}\langle \mathcal{P}^{{\rm Adj}}_{\Omega}(\bm{V}), \bm{a}(\theta_k)\bm{\varphi}^H_k\bm{E}_k\rangle\nonumber\\
&=\sup_{\substack{\theta_k\in (0,\pi),\bm{\varphi}_k, \\\|\bm{E}_k\|_{2\rightarrow 2}\le C_e}}\|(\mathcal{P}^{{\rm Adj}}_{\Omega}(\bm{V}))^H\bm{a}(\theta_k)\|_2\|\bm{E}_k^{\mathsf{H}}\bm{\varphi}_k\|_2, \nonumber
\end{align}
where in the last step above, we used H\"{o}lder's inequality. A sufficient condition for the latter constraints in \eqref{d4} to hold is given in the following relation: 
\begin{align}\label{eq.const1}
	&	\|(\mathcal{P}^{{\rm Adj}}_{\Omega}(\bm{V}))^{\mathsf{H}}\bm{a}(\theta_k)\|_2 C_e \max_{k}\|\bm{\phi}_k\|_2\le \min_k \beta_k, 
\end{align} 
for $\theta_k \in (0,\pi)$, where we used the unitary properties of DFT and permutation matrices and that $\|\bm{g}\|_{\infty}\le \zeta$. Then, by defining the notation $c_1$ and \eqref{eq.atoms}, \eqref{eq.const1} can be rewritten as follows:
\begin{align}\label{pl1}
\sum_{i=1}^T \Big|\sum_{l=1}^N (\mathcal{P}^{{\rm Adj}}_{\Omega}(\bm{V}))_{(l,i)} c_1 {\rm e}^{-j 2\pi (l-1) \cos(\theta)}\Big|^2\le 1,
\end{align}
where $ \theta\in (0,\pi)$. Next, we define $\bm{a}_i$ as $i$-th column of  $\mathcal{P}^{{\rm Adj}}_{\Omega}(\bm{V})$, i.e., 
$\bm{a}_i:=c_1[(\mathcal{P}^{{\rm Adj}}_{\Omega}(\bm{V}))_{(1,i)},..., (\mathcal{P}^{{\rm Adj}}_{\Omega}(\bm{V}))_{(N,i)}]^{\mathsf{T}}$ and $\bm{f}(\theta):=[1,..., {\rm e}^{j 2\pi (N-1) \cos(\theta)}]^{\mathsf{T}}$, \eqref{pl1} can be reformulated as
\begin{align}
	\sum_{i=1}^T|\bm{f}(\theta)^{\mathsf{H}}\bm{a}_i|^2\leq 1,
\end{align} 
which implies that the polynomial $1-\bm{f}(\theta)^{\mathsf{H}}\sum_{i=1}^T\bm{a}_i\bm{a}_i^{\mathsf{H}}\bm{f}(\theta)$ is non-negative. Based on \cite[Theorem 1.1]{dumitrescu2017positive}, there exists a polynomial $\bm{q}_1(\theta):=\sum_{i=1}^N b_i{\rm e}^{-j 2\pi (i-1) \cos(\theta)}= \bm{f}(\theta)^{\mathsf{H}}\bm{b}$ for some $\bm{b}\in\mathbb{C}^{N}$ that
\begin{align}\label{pl2}
	1-\bm{f}(\theta)^{\mathsf{H}}\sum_{i=1}^T&\bm{a}_i\bm{a}_i^{\mathsf{H}}\bm{f}(\theta)=|q_1(\theta)|^2
 =\bm{f}(\theta)^{\mathsf{H}}\bm{b}\bm{b}^{\mathsf{H}}\bm{f}(\theta).
\end{align}
Set $\bm{Q}=\sum_{i=1}^T\bm{a}_i\bm{a}_i^{\mathsf{H}}+\bm{b}\bm{b}^{\mathsf{H}}$. Since both $\bm{Q}$ and $\bm{Q}-\sum_{i=1}^T\bm{a}_i\bm{a}_i^{\mathsf{H}}$ are positive semidefinite, using Schur complement lemma \cite[Appendix A.5.5]{boyd2004convex} leads to the semidefinite constraint:
\begin{align}
    \label{eq:Schur}
	\begin{bmatrix}
		\bm{Q}& \mathcal{P}^{{\rm Adj}}_{\Omega}(\bm{V}) c_1\\
		(\mathcal{P}^{{\rm Adj}}_{\Omega}(\bm{V}))^{\mathsf{H}} c_1&\bm{I}_T
	\end{bmatrix}\succeq \bm{0}. 
\end{align} 
Further, by \eqref{pl2}, we have:
\begin{align}
    \label{eq:Polyf}
	\bm{f}^{\mathsf{H}}(\theta)\bm{Q}\bm{f}(\theta)=\langle \bm{Q}, \bm{f}(\theta)\bm{f}^{\mathsf{H}}(\theta) \rangle=1, \forall \theta \in (0,\pi).
\end{align} 
With the same arguments as in  \cite[Appendix C]{SajadBOD2023} (see Eq. (51) to Eq (67)), we can show that   \eqref{eq:Polyf} leads to 
\begin{align}
    \label{eq:Qconstraint}
    \langle \mathcal{T}(\bm{e}_q), \bm{Q}  \rangle=1_{q=0}, q=-N+1,..., N-1.
\end{align}
Therefore, \eqref{eq:Qconstraint} and \eqref{eq:Schur} together  lead to $\sum_{i=1}^T|\bm{f}(\theta)^{\mathsf{H}}\bm{a}_i|^2\le 1$ which is equivalent to \eqref{d4}.

\bibliographystyle{ieeetr}
\bibliography{Ref}

\begin{thebibliography}{10}

\bibitem{SaeedBlind2022}
S.~Razavikia, J.~A. Peris, J.~M.~B. Da~Silva, and C.~Fischione, ``Blind
  asynchronous over-the-air federated edge learning,'' in {\em IEEE Globecom
  Workshops}, pp.~1834--1839, 2022.

\bibitem{mahmood2020six}
N.~H. Mahmood, H.~Alves, O.~A. L{\'o}pez, M.~Shehab, D.~P.~M. Osorio, and
  M.~Latva-Aho, ``Six key features of machine type communication in {6G},'' in
  {\em 6G Wireless Summit}, pp.~1--5, IEEE, 2020.

\bibitem{razavikia2023computing}
S.~Razavikia, J.~M. B.~d. Silva~Jr, and C.~Fischione, ``Computing functions
  over-the-air using digital modulations,'' in {\em IEEE Inter. Conf. on
  Commun.}, 2023.

\bibitem{polyanskiy2017perspective}
Y.~Polyanskiy, ``A perspective on massive random-access,'' in {\em IEEE Inter.
  Symp. on Info. Theo.}, pp.~2523--2527, 2017.

\bibitem{hasan2013random}
M.~Hasan, E.~Hossain, and D.~Niyato, ``Random access for machine-to-machine
  communication in {LTE}-advanced networks: Issues and approaches,'' {\em IEEE
  Commun. Mag.}, vol.~51, no.~6, pp.~86--93, 2013.

\bibitem{bjornson2017random}
E.~Bj{\"o}rnson, E.~De~Carvalho, J.~H. S{\o}rensen, E.~G. Larsson, and
  P.~Popovski, ``A random access protocol for pilot allocation in crowded
  massive {MIMO} systems,'' {\em IEEE Trans. on Wire. Commun.}, vol.~16, no.~4,
  pp.~2220--2234, 2017.

\bibitem{chen2021sparse}
Z.~Chen, F.~Sohrabi, and W.~Yu, ``Sparse activity detection in multi-cell
  massive {MIMO} exploiting channel large-scale fading,'' {\em IEEE Trans. on
  Sig. Proc.}, 2021.

\bibitem{fengler2022pilot}
A.~Fengler, O.~Musa, P.~Jung, and G.~Caire, ``Pilot-based unsourced random
  access with a massive {MIMO} receiver, interference cancellation, and power
  control,'' {\em IEEE Jour. on Sel. Areas in Commun.}, 2022.

\bibitem{xie2022massive}
X.~Xie, Y.~Wu, J.~An, J.~Gao, W.~Zhang, C.~Xing, K.-K. Wong, and C.~Xiao,
  ``Massive unsourced random access: Exploiting angular domain sparsity,'' {\em
  IEEE Trans. on Commun.}, 2022.

\bibitem{liu2018sparse}
L.~Liu, E.~G. Larsson, W.~Yu, P.~Popovski, C.~Stefanovic, and E.~De~Carvalho,
  ``Sparse signal processing for grant-free massive connectivity: A future
  paradigm for random access protocols in the internet of things,'' {\em IEEE
  Sig. Proc. Mag.}, vol.~35, no.~5, pp.~88--99, 2018.

\bibitem{SajadBOD2023}
S.~Daei and M.~Kountouris, ``Blind goal-oriented massive access for future
  wireless networks,'' {\em IEEE Trans. on Sig. Proc.}, pp.~1--16, 2023.

\bibitem{daei2022Random}
A.~Afshar, V.~Tabataba~Vakili, and S.~Daei, ``Active user detection and channel
  estimation for spatial-based random access in crowded massive {MIMO} systems
  via blind super-resolution,'' {\em IEEE Sig. Proc. Letters}, 2022.

\bibitem{ke2020compressive}
M.~Ke, Z.~Gao, Y.~Wu, X.~Gao, and R.~Schober, ``Compressive sensing based
  adaptive active user detection and channel estimation: Massive access meets
  massive {MIMO},'' {\em IEEE Trans. on Sig. Proc.}, 2020.

\bibitem{liu2018massive}
L.~Liu and W.~Yu, ``Massive connectivity with massive {MIMO}—part i: Device
  activity detection and channel estimation,'' {\em IEEE Trans. on Sig. Proc.},
  vol.~66, no.~11, pp.~2933--2946, 2018.

\bibitem{ding2019analysis}
J.~Ding, D.~Qu, and J.~Choi, ``Analysis of non-orthogonal sequences for
  grant-free {RA} with massive {MIMO},'' {\em IEEE Trans. on Commun.}, vol.~68,
  no.~1, pp.~150--160, 2019.

\bibitem{fengler2021non}
A.~Fengler, S.~Haghighatshoar, P.~Jung, and G.~Caire, ``Non-bayesian activity
  detection, large-scale fading coefficient estimation, and unsourced random
  access with a massive {MIMO} receiver,'' {\em IEEE Trans. on Info. Theo.},
  vol.~67, no.~5, pp.~2925--2951, 2021.

\bibitem{bjornson2016random}
E.~Bj{\"o}rnson, E.~De~Carvalho, E.~G. Larsson, and P.~Popovski, ``Random
  access protocol for massive {MIMO}: Strongest-user collision resolution,'' in
  {\em IEEE Inter. Conf. on Commun.}, pp.~1--6, 2016.

\bibitem{djelouat2021joint}
H.~Djelouat, M.~Leinonen, L.~Ribeiro, and M.~Juntti, ``Joint user
  identification and channel estimation via exploiting spatial channel
  covariance in {mMTC},'' {\em IEEE Wire. Commun. Letters}, vol.~10, no.~4,
  pp.~887--891, 2021.

\bibitem{ma2018sparse}
J.~Ma, S.~Zhang, H.~Li, F.~Gao, and S.~Jin, ``Sparse bayesian learning for the
  time-varying massive {MIMO} channels: Acquisition and tracking,'' {\em IEEE
  Trans. on Commun.}, vol.~67, no.~3, pp.~1925--1938, 2018.

\bibitem{zhang2017blind}
J.~Zhang, X.~Yuan, and Y.-J.~A. Zhang, ``Blind signal detection in massive
  {MIMO}: Exploiting the channel sparsity,'' {\em IEEE Trans. on Commun.},
  vol.~66, no.~2, pp.~700--712, 2017.

\bibitem{sayyari2020blind}
S.~Sayyari, S.~Daei, and F.~Haddadi, ``Blind two-dimensional super-resolution
  in multiple-input single-output linear systems,'' {\em IEEE Sig. Proc.
  Letters}, 2020.

\bibitem{Seidi2022Novel}
M.~Seidi, S.~Razavikia, S.~Daei, and J.~Oberhammer, ``A novel demixing
  algorithm for joint target detection and impulsive noise suppression,'' {\em
  IEEE Commun. Letters}, vol.~26, no.~11, pp.~2750--2754, 2022.

\bibitem{SaeedBinary2020}
S.~Razavikia, A.~Amini, and S.~Daei, ``Reconstruction of binary shapes from
  blurred images via {Hankel}-structured low-rank matrix recovery,'' {\em IEEE
  Trans. on Image Proc.}, vol.~29, pp.~2452--2462, 2020.

\bibitem{daei2019living}
S.~Daei, F.~Haddadi, and A.~Amini, ``Living near the edge: A lower-bound on the
  phase transition of total variation minimization,'' {\em IEEE Trans. on Info.
  Theo.}, vol.~66, no.~5, pp.~3261--3267, 2019.

\bibitem{daei2019error}
S.~Daei, F.~Haddadi, A.~Amini, and M.~Lotz, ``On the error in phase transition
  computations for compressed sensing,'' {\em IEEE Trans. on Info. Theo.},
  vol.~65, no.~10, pp.~6620--6632, 2019.

\bibitem{daei2019exploiting}
S.~Daei, F.~Haddadi, and A.~Amini, ``Exploiting prior information in
  block-sparse signals,'' {\em IEEE Trans. on Sig. Proc.}, vol.~67, no.~19,
  pp.~5093--5102, 2019.

\bibitem{maskan2023demixing}
H.~Maskan, S.~Daei, and M.~H. Kahaei, ``Demixing sines and spikes using
  multiple measurement vectors,'' {\em Signal Processing}, vol.~203, p.~108786,
  2023.

\bibitem{valiulahi2019two}
I.~Valiulahi, S.~Daei, F.~Haddadi, and F.~Parvaresh, ``Two-dimensional
  super-resolution via convex relaxation,'' {\em IEEE Trans. on Sig. Proc.},
  vol.~67, no.~13, pp.~3372--3382, 2019.

\bibitem{candes2014towards}
E.~J. Cand{\`e}s and C.~Fernandez-Granda, ``Towards a mathematical theory of
  super-resolution,'' {\em Communications on pure and applied Mathematics},
  vol.~67, no.~6, pp.~906--956, 2014.

\bibitem{tang2013compressed}
G.~Tang, B.~N. Bhaskar, P.~Shah, and B.~Recht, ``Compressed sensing off the
  grid,'' {\em IEEE Trans. on Info. Theo.}, vol.~59, no.~11, pp.~7465--7490,
  2013.

\bibitem{fernandez2016super}
C.~Fernandez-Granda, ``Super-resolution of point sources via convex
  programming,'' {\em Information and Inference: A Journal of the IMA}, vol.~5,
  no.~3, pp.~251--303, 2016.

\bibitem{grant2014cvx}
M.~Grant and S.~Boyd, ``{CVX}: {MATLAB} software for disciplined convex
  programming, version 2.1,'' 2014.

\bibitem{dumitrescu2017positive}
B.~Dumitrescu, {\em Positive Trigonometric Polynomials and Signal Processing
  Applications}.
\newblock Springer, 2017.

\bibitem{boyd2004convex}
S.~P. Boyd and L.~Vandenberghe, {\em Convex optimization}.
\newblock Cambridge university press, 2004.

\end{thebibliography}
\end{document}